\documentclass[conference]{IEEEtran}
\IEEEoverridecommandlockouts
\usepackage{booktabs}
\usepackage[numbers]{natbib}
\usepackage{amsmath,amssymb,amsfonts}
\usepackage{algorithmic}
\usepackage{graphicx}
\usepackage{textcomp}
\usepackage{xcolor}
\usepackage{subcaption}
\usepackage{multirow}

  \makeatletter
  \DeclareRobustCommand{\cev}[1]{%
    {\mathpalette\do@cev{#1}}%
  }
  \newcommand{\do@cev}[2]{%
    \vbox{\offinterlineskip
      \sbox\z@{$\m@th#1 x$}%
      \ialign{##\cr
        \hidewidth\reflectbox{$\m@th#1\vec{}\mkern4mu$}\hidewidth\cr
        \noalign{\kern-\ht\z@}
        $\m@th#1#2$\cr
      }%
    }%
  }
  \makeatother
  \newcommand{\vect}[1]{\boldsymbol{#1}}

\def\BibTeX{{\rm B\kern-.05em{\sc i\kern-.025em b}\kern-.08em
    T\kern-.1667em\lower.7ex\hbox{E}\kern-.125emX}}

    \usepackage{array}

    \newcolumntype{P}[1]{>{\centering\arraybackslash}p{#1}}

    \usepackage{hyperref}  

\begin{document}

\title{Host Load Prediction with Bi-directional Long Short-Term Memory in Cloud Computing*\\
}

\author{
\IEEEauthorblockN{1\textsuperscript{st} Hengheng Shen}
\IEEEauthorblockA{
\textit{Institute of Computing Technology, Chinese Academy}\\
Beijing, China \\
shenhengheng17g@ict.ac.cn}
\and
\and
\and
\IEEEauthorblockN{1\textsuperscript{st} Xuehai Hong}
\IEEEauthorblockA{
\textit{Institute of Computing Technology, Chinese Academy}\\
Beijing, China \\
hxh@ict.ac.cn}
}

\maketitle

\begin{abstract}
    Host load prediction is the basic decision information for managing 
    the computing resources usage on the cloud platform, its accuracy is  critical 
    for achieving the service-level agreement.
    Host load data in cloud environment is more high volatility and noise compared to that of  grid computing, 
    traditional data-driven methods tend to have low predictive accuracy when dealing with  host load 
    of cloud computing,
    Thus, we have proposed a host load prediction method  based on Bi-directional Long Short-Term Memory (BiLSTM) in this paper. 
   Our BiLSTM-based apporach improve the memory capbility and nonlinear modeling ability of LSTM and LSTM Encoder-Decoder (LSTM-ED), which is used in the recent previous work,
   In order to evaluate our approach, we have conducted experiments using a 1-month trace of a 
   Google data centre with more than twelve thousand machines.
   our BiLSTM-based approach successfully achieves higher accuracy than other 
   previous models, including the recent LSTM one and LSTM-ED one.

\end{abstract}


\section{Introduction}\label{sec:1}

Cloud computing is a resource service model, computing model, and on-demand billing model 
that provides elastic and scalable virtualization through the Internet as a service, 
but the widespread use of cloud computing has led to an increasing cost of enterprise 
investment in data centers, which note that the cost of data centers accounts for about 25\% of the total budget
of the enterprise's IT business. Although the data center has a large number of users and market 
share dominates, inefficient data centers can cause them to fail. One of the main causes of
data center inefficiencies is low server utilization. Gartner and McKinsey report 
that server utilization in most enterprise data centers is only 6\%--12\%~\cite{shehabi2016united},
and  Amazon AWS server utilization is only 7\%--17\%~\cite{6118751}.  
 To address the problem of underutilization of data center servers, we need to migrate 
virtual machines based on the load of the host. In this process, how to accurately 
predict the load of each physical host is a key issue. If you can accurately predict 
the load change trend for each physical host before scheduling, and advance the migration
of virtual machines and resource scheduling to avoid violating service-level agreement (SLA),
it will ultimately benefit the load balancing of the entire cloud platform.

Most previous works\cite{6212065,10.5555/1862805.1862814,1336711,6008748,1213129,8457781,article} 
on host load prediction have focused on the host load in 
traditional Grids. However, unlike the applications used in a Grid, 
the tasks in a Cloud tend to be shorter and more interactive. 
According to the comparison of work load set ween Cloud and Grid~\cite{6337784,di2012host}, 
the average noise in a Cloud is approximately 20 times larger than the average noise in a Grid. 
Therefore, predicting the host load in a Cloud is more diffcult than in
 a Grid. Specifically, previous methods achieve limited accuracy when 
 they are applied to the cloud environment.

In this paper, we predict the host load multi-step-ahead with a model called  
Bi-directional
long short-term memory (BiLSTM) which is concise yet adaptive and powerful. 
During host load prediction, the quantity of history information required to 
predict future values may be variant in different load traces. Unlike previous methods, 
the BiLSTM model can learn how long does it really need instead of a manually control. 
Furthermore, this method is an end-to-end model which never requires an extra 
feature-extracted step. Experiment results show that our method achieves better 
performance than other start-of-the-art methods.

The outline of the paper is as follows. Section~\ref{sec:2} gives an overview of the 
related work and comparisons between each other. The architecture of our 
proposed method is shown in Section~\ref{sec:3}. In Sect. \ref{sec:4}, we present 
the experiment results and comparisons. Finally, we conclude 
the paper and future work in Section~\ref{sec:5}.

\section{Related work}\label{sec:2}

Host load prediction in grid and cloud systems has 
gathered a lot of attention from researchers
 due to its benefits in improving resource allocation and utilization 
 while satisfying service-level agreement (SLA).

 Many efforts have been made toward host load prediction in Grids or HPC systems.
 Khan et al.~\cite{6212065} used the hidden Markov model to establish a
  CPU load prediction model based 
 on self-similarity for the CPU load of the cloud data center. Dabrowski et al.~\cite{10.5555/1862805.1862814} 
 used Markov model modeling to predict host load data in a simulated 
 cloud environment. 
 Akioka et al.~\cite{1336711} proposed a load prediction framework by combining 
 Gaussian Hidden Markov Model and seasonal variance analysis for the host 
 load of the grid system, predicted the host load value of the grid system 
 at the next moment. Firstly, the clustering method based on partitioning the underlying bipartite graph
  is used to capture the workload similarity characteristics of different 
 groups of virtual machines, then the hidden Markov model is used to model the time 
correlation, finally the different clusters 
are predicted. 
For the workload of each virtual machine, if the markov chain 
stochastic model works normally, on the one hand, the random process needs 
to have no memory, on the other hand, the system involved is stable. Obviously, 
the cloud computing resource allocation is dynamically changing, so the markov 
chain stochastic model can only be applied to short-term actual load prediction.
Roy et al.~\cite{6008748} predicted the workload in the cloud computing environment based 
on the Exponential Moving Average (EMA) method. 
Yang et al.~\cite{1213129} proposed a 
prediction method for the CPU usage of the host, using the stability and 
trend assumptions of the time series,
 combined with the prediction error of the last time step and 
 the observation data to dynamically adjust the parameters of the model. 
 The experiment shows that the accuracy of the CPU load prediction obtained 
 by the proposed trend seasonal prediction method is significantly higher 
 than other linear-based prediction method.
Kim et al.~\cite{8457781} proposed a combination 
 method for cloud data center workload prediction which combining models such 
 as ARMA, linear regression, and support vector machines, and using regression 
 algorithms to dynamically determine the weight of each predictor, experiments 
 show that the method can accurately predict the host load and 
 workload of the data center.

 The artificial neural networks are widely used on predict host load of  grid and HPC systems, 
 host load prediction research due to their high nonlinearity, adaptability and arbitrary 
 function fitting. In addition, with the further advancement of deep learning, methods based 
 on artificial neural networks have received unprecedented attention in academia and industry. 
 Feedforward Neural Network (FNN) is a type of traditional neural network widely used in host 
 load prediction, in which historical host load data is used for FNN input, and the host 
 load value is directly used as the output of FNN, and iteratively adjusts its internal 
 parameter values based on a specific optimization algorithm to model and analyze the highly 
 nonlinear relationship between input and output. 
 Prevost et al.~\cite{5966610} took the resource usage 
 data of the cloud host as input, the workload resource usage of the cloud data center application as 
 output, and trained the FNN based on the back propagation algorithm, and finally used the 
 trained model for online workload prediction. 
 Duy et al.~\cite{5160878} used the length of the 
 ready queue maintained by the scheduler in the grid computing system as the CPU load 
 index, and built a three-layer FNN to predict the CPU load in the future.
 Yang et al.~\cite{yang2013host} proposed a new method for the host load prediction problem of 
 Google Cloud Data Center. This method combines the phase space reconstruction (PSR) 
 method and the grouping data processing method based on evolutionary algorithm (EA-GMDH). 
 The host load in the distributed environment and the host load in the cloud computing environment 
 are predicted separately. Considering that the host load is a single variable time series, 
 the PSR method is used to reconstruct the one-dimensional host load sequence in the 
 multi-dimensional phase space, and the EA-GMDH method optimizes the model 
 parameters of the FNN to select the best model. 
 However, this method is limited by the 
 number of neurons in the multi-step prediction task in advance and the FNN cannot learn 
 long-term dependencies. so the prediction accuracy is poor. 
 In order to solve the problem 
 of multi-step early prediction of host load, Yang et al.~\cite{yang2014new} proposed to use an 
 autoencoder as a feature extraction layer and an echo state network (ESN) as a 
 prediction network to achieve multi-step early host load prediction. 
 Since the random and sparse dynamic storage layer of ESN is used as the 
 non-linear feature extraction layer and the input data storage layer, the accuracy 
 is improved compared to the PSR+EA-GMDH model, but the ESN model uses manual 
 selection of parameters, it reduces  generalization ability of different 
 loads, and ESN relies heavily on the features extracted from the auto-encoder, 
 which adds additional workload for model parameter adjustment, so ESN is 
 less effective in the actual cloud host load multi-step prediction task in advance.

Song et al.~\cite{song2018host} used the advantages of long short-term memory 
in the recurrent neural network to model time series data, 
and realized the time-correlation modeling and multi-step advancement of 
the host load  prediction in the cloud computing environment and grid computing system.
Peng et al.~\cite{peng2018multi} apply the gated recurrent units  
  to cloud computing host load prediction, and adopted the encoder-decoder network 
 (GRUED) based on the gated recurrent units as the load prediction network, 
 using traditional dataset `Dinda' of grid computing system and Google cluster trace dataset 
 to verify. The experimental results show that the GRUED model performs better than 
 the prediction model based on long  short-term memory on the two datasets.
 Nguyen et al.~\cite{nguyen2019host} proposed a composite deep neural network structure model based on 
 long short-term memory encoder-decoder network (LSTM-ED) and FNN  which 
 normalized and serialized host load data used as the input of the proposed model, 
 LSMT-ED network is used to achieve deep feature extraction of the input data, 
 FNN network is used to further complete the regression analysis  of the 
 extracted features and host load, and the proposed method in the Google cluster 
 trace data has been successfully applied.


\section{Our Method}\label{sec:3}

\subsection{Overview of proposed method}

\begin{figure}[htbp]
    \centerline{\includegraphics[width=0.45\textwidth]{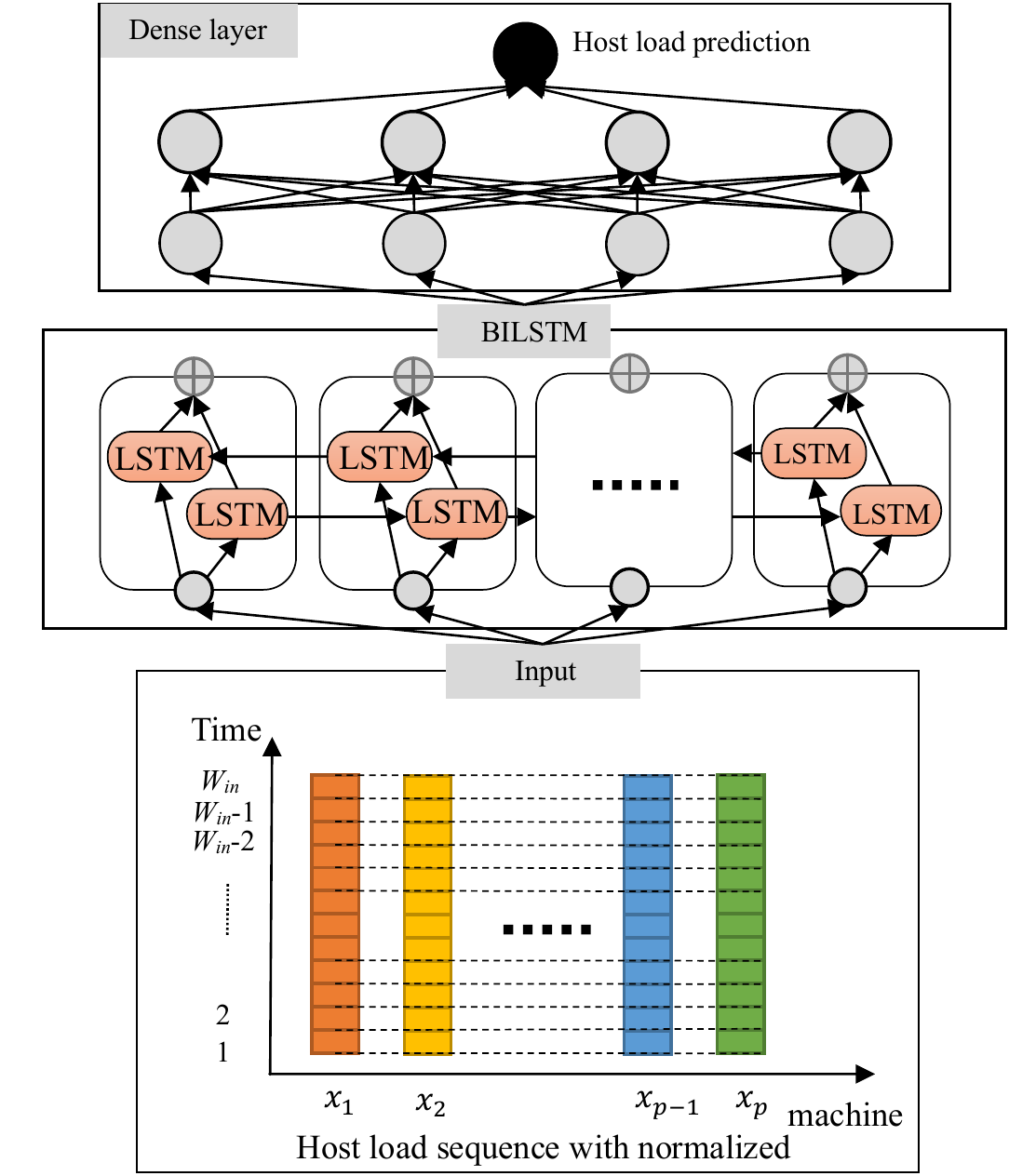}}
    \caption{The architecture of our proposed method}
    \label{fig:arch}
    \end{figure}

Figure.~\ref{fig:arch} presents the architecture overview of our proposed model.
At the center of 
our mothods, we use a model called Bi-directional Long Short-Term Memory (BiLSTM) for 
host load prediction.
The BiLSTM consists of two main components, including a
forward LSTM computinng procedure which using historical host load to update hidden state with forwarding,
 and a 
backward LSTM computinng procedure which using future host load to update hidden state with backwarding.

It can be seen from the Figure.~\ref{fig:arch} that the time-correlation features hidden 
in the input sequence data are first extracted. 
The extraction process is implemented by the BiLSTM network. 
Specifically, the BiLSTM network will perform feature extraction 
on the long-term dependence of the input sequence layer
 by layer and adopt the forward and backward calculation processes respectively;
  secondly, the extracted features are further feed into the fully connected layer;
   finally the future host load is predicted through a linear regression output layer.

    The host load time series is divided into consecutive `history' sequences of fixed size; 
    each of this `history' sequence is accompanied by a `prediction' sequence of fixed size. 
    The ‘history’ and ‘prediction’ sequences are used as inputs and supervised outputs/labels
     for the BiLSTM, respectively. Depending on the performed tasks, these `prediction'
     sequences can be either real host load values or mean load values over future time intervals.

     In addition, the Google load trace provides measurements taken at 1-s intervals, 
     which is too small compared to the usual CPU load fluctuations in the trace. 
     Thus, in this work the smallest interval we have used to take samples
      from the trace is 5 min, which is the same as previous works that used this case for comparison purpose.

\subsection{Recurrent neural network}\label{rnn}

\begin{figure}[htbp]
    \centerline{\includegraphics[width=0.4\textwidth]{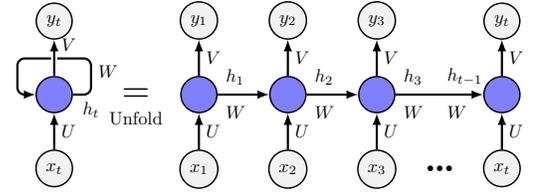}}
    \caption{Recurrent Neural Network and its unfold structure.}
    \label{fig:rnn}
    \end{figure}

Recurrent Neural Network is a type of network with loops in them to allow information to persist. 
This type of network can be used for our host load prediction task by predicting the next value 
based on historical load values. Figure \ref{fig:rnn} shows the architecture of a Recurrent Neural Network (RNN) 
with one hidden layer together with its unrolled form, 
in which at time step $t$ in the network $x_t$ is the input, $h_t$ is the hidden state, 
and $y_t$ is the output.

In our host load prediction task, $x_t$ can be the historical load value (possibly after normalization). 
Then the hidden state s t of RNN can be calculated based on the previous hidden state and the output at the current time step:
    \begin{equation}
        \vect{h}^t=\sigma \left( \vect{U} \vect{x}^t+ \vect{W} \vect{h}^{t-1}+ \vect{b}_h   \right)
    \end{equation}
    
    where $\sigma$ is usually a nonlinear function like \verb|tanh| 
    or 
    \verb|ReLU|.
    To calculate the first hidden state, $\vect{h}_{-1}$ is typically initialized to zeros.
    
    The output state $\vect{y}_t$ can be calculated based on the hidden state $\vect{h}_t$ as follows:
    \begin{equation}
        \hat{\vect{y}}^{t}=\vect{V} \vect{h}^t+ \vect{b}_y
    \end{equation}

    Different from a traditional deep neural network, RNN uses the same set of parameters
     $\left(\vect{U}, \vect{V}, \vect{W} \right)$ across all steps, 
     which greatly reduces the number of parameters the network needs to learn.

\subsection{Long short-term memory}\label{lstm}

Long Short-Term Memory (LSTM) is a special kind of RNN, which can resolve the vanishing gradients issue and 
is capable of learning long-term dependencies. Introduced by Hochreiter et al. \cite{hochreiter1997long} in 1997, 
there have been many works which apply LSTM. These include the recent previous work by Song et al. \cite{song2018host}, 
where they optimized the parameters and showed that their LSTM model outperformed other previous models in the Google load trace.

Figure \ref{fig:lstm} shows that architecture overview of a LSTM cell, 
which consists of the following components:

    \begin{figure}[htbp]
        \centerline{\includegraphics[width=0.4\textwidth]{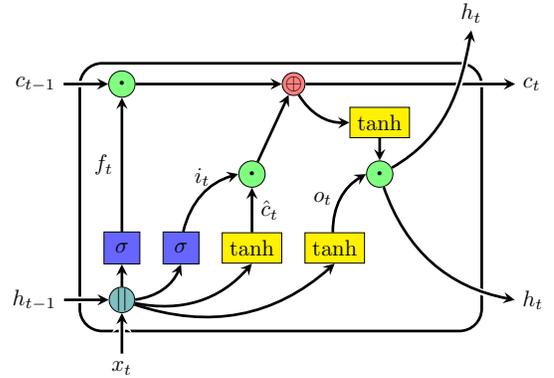}}
        \caption{The LSTM memory block.}
        \label{fig:lstm}
        \end{figure}
        
 \begin{itemize}
     \item $x_t$: external input at time step $t$.
     \item $h_{t-1}$: hidden state at times $\left( t- 1\right)$ or $t$. This is also used as output or input for 
     the next layer of LSTM cells (in multi-layer LSTM).
     \item $c_{t-1}$, $c_t$: the `cell state' or `memory' at time step $t-1$ or $t$.
     \item $f_t$: the result of the forget gate, which controls whether to forget (for values close to zero) or 
     remember (for values close to one) the memory $c_{t-1}$.
     \item $i_t$: the result of the input gate, which determines the degree of importance of 
     the (transformed) new external input.
     \item $\hat{c}_t$: the result of the candidate cell state, which performs a nonlinear transformation of the new external input $x_t$ .
     \item $o_t$: the result of the output gate, which controls the amount of the new cell state $c_t$
     that goes to the output and the hidden state. 
 \end{itemize}

 Every time the LSTM takes an input $x_t$, the three gates $f_t$, $i_t$, $o_t$ and the candidate cell state $c_t$ are updated as follows:
 \begin{equation}
    \begin{aligned}
        \vect{i}_t &= \sigma\left(\vect{W}_u \left[ \vect{h}_{t-1}; \vect{x}_{t} \right]+\vect{b}_u\right)\\
        \vect{f}_t &= \sigma\left(\vect{W}_f \left[ \vect{h}_{t-1}; \vect{x}_{t} \right]+\vect{b}_f\right)\\
        \vect{o}_t &= \sigma\left(\vect{W}_o \left[ \vect{h}_{t-1}; \vect{x}_{t} \right]+\vect{b}_o\right)\\
        \hat{\vect{c}}_{t} &= \tanh\left(\vect{W}_c\left[\vect{h}_{t-1}; \vect{x}_{t} \right] + \vect{b}_c\right)\\
    \end{aligned}
\end{equation}

where we use sigmoid function for $\sigma$ in this work.

The cell state at the current time step can be updated using the results from 
the cell gate and the cell state at the last 
time step as follows:
\begin{equation}
    \begin{aligned}
        \vect{c}_{t} &=  \vect{f}_t  \odot  \vect{c}_{t-1} +  \vect{i}_t \odot \hat{\vect{c}}_{t} \\
    \end{aligned}
\end{equation}

Finally, the hidden state (or output) can be updated using 
the results from the output gate and the cell state at the current time step as follows:
\begin{equation}
        \vect{h}_{t} = \vect{o}_t \odot \tanh( \vect{c}_{t})
\end{equation}

\subsection{Bi-directional long short-term memory}\label{BiLSTM}

\begin{figure}[htbp]
    \centerline{\includegraphics[width=0.45\textwidth]{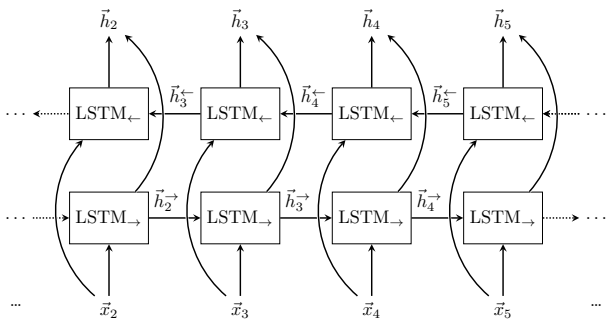}}
    \caption{A bidirectional LSTM network.}
    \label{fig:bilstm}
    \end{figure}

In order to further improve upon the previous work’s LSTM model that shows longterm 
dependencies learning capability, in this work we have used a model called Bi-directional Long 
Short-Term Memory (BiLSTM). 
The BiLSTM model consists of two directional LSTM RNN units that act as an forward and backward pair, 
as illustrated in Figure.~\ref{fig:bilstm}.

The BiLSTM-based network used  is based on the basic LSTM unit. 
Section~\ref{lstm} has briefly introduced the internal operation mechanism of the LSTM unit. 
Essentially. The calculation 
process of the forward LSTM unit is described, that is, the history hidden state information 
$\vect{h}_{t-1}$ and the current input information $\vect{x}_t$ are combined to realize the update and output 
of the hidden state information $\vect{h}_t$ at the current time. The above calculation process 
of the forward LSTM unit, It can be summarized as follows:
\begin{equation}
    \begin{aligned}
         \vec {\vect{h}}_{t} &=  f_1\left(\vec{\vect{x}}_t; \vec{\vect{h}}_{t-1}; \vec{\vect{\Theta}}_{\rm LSTM} \right)  \\
           &= \left\{
\begin{aligned}
    \vec{\vect{i}}_t &= \sigma\left(\vec{\vect{W}}_u \left[ \vec{\vect{h}}_{t-1}; \vec{\vect{x}}_{t} \right]+\vec{\vect{b}}_u\right)\\
    \vec{\vect{f}}_t &= \sigma\left(\vec{\vect{W}}_f \left[ \vec{\vect{h}}_{t-1}; \vec{\vect{x}}_{t} \right]+\vec{\vect{b}}_f\right)\\
    \vec {\vect{o}}_t &= \sigma\left(\vec{\vect{W}}_o \left[ \vec{\vect{h}}_{t-1}; \vec{\vect{x}}_{t} \right]+\vec{\vect{b}}_o\right)\\
    \vec {\hat{\vect{c}}}_{t} &= \tanh\left(\vec {\vect{W}}_c\left[\vec{\vect{h}}_{t-1}; \vec{\vect{x}}_{t} \right] +\vec{\vect{b}}_c\right)\\
    \vec {\vect{c}}_{t} &=  \vec{\vect{f}}_t  \odot \vec {\vect{c}}_{t-1} +  \vec {\vect{i}}_{t} \odot \vec {\hat{\vect{c}}}_{t} \\
    \vec {\vect{h}}_{t} &= \vect{o}_t \odot \tanh( \vect{c}_{t})
\end{aligned}
\right.
    \end{aligned}
    \label{eq:1}
\end{equation}

The core idea of the bidirectional LSTM unit is to use two separate hidden layers to 
process sequence data from the forward and backward directions 
to model the impact of historical and future information 
on the current hidden state, respectively.
 Equation~(\ref{eq:1}) describes the internal calculation process of the forward LSTM unit. 
The reverse LSTM unit mainly uses the future information $\vect{h}_{t+1}$ to update the hidden state 
information, as shown in the following formula:
\begin{equation}
    \begin{aligned}
        \cev {\vect{h}}_{t} &= f_1\left(\cev{\vect{x}}_t; \cev{\vect{h}}_{t+1}; \cev{\vect{\Theta}}_{\rm LSTM} \right)\\
                            &= \left\{
\begin{aligned}
    \cev{\vect{i}}_t &= \sigma\left(\cev{\vect{W}}_u \left[ \cev{\vect{h}}_{t+1}; \cev{\vect{x}}_{t} \right]+\cev{\vect{b}}_u\right)\\
    \cev{\vect{f}}_t &= \sigma\left(\cev{\vect{W}}_f \left[ \cev{\vect{h}}_{t+1}; \cev{\vect{x}}_{t} \right]+\cev{\vect{b}}_f\right)\\
    \cev {\vect{o}}_t &= \sigma\left(\cev{\vect{W}}_o \left[ \cev{\vect{h}}_{t+1}; \cev{\vect{x}}_{t} \right]+\cev{\vect{b}}_o\right)\\
    \cev{\hat{\vect{c}}}_{t} &= \tanh\left(\cev {\vect{W}}_c\left[\cev{\vect{h}}_{t+1}; \cev{\vect{x}}_{t} \right] +\cev{\vect{b}}_c\right)\\
    \cev {\vect{c}}_{t} &=  \cev{\vect{f}}_t  \odot \cev {\vect{c}}_{t+1} +  \cev {\vect{i}}_{t} \odot \cev {\hat{\vect{c}}}_{t} \\
    \cev {\vect{h}}_{t} &= \vect{o}_t \odot \tanh( \vect{c}_{t})
\end{aligned}
\right.
    \end{aligned}
    \label{eq:2}
\end{equation}

In summary, it can be seen from Figure~\ref{fig:arch} that the host load prediction model based 
on our proposed method mainly includes three steps, 
and the technical details of the model are now described as follows:

\paragraph{Step 1} Feed the input data $X_i=\left[x_1, \cdots, x_{W_{in}}\right]$ into 
the BiLSTM network to extract the temporal features of the  host load. 
The output features can be calculated as follows:
\begin{equation}
    \vect{H}_i=[\vect{h}_1,\cdots,\vect{h}_{W_{in}} ]=f_1 (\vect{X}_i;\vec{\vect{\Theta}}_{\rm LSTM},\cev{\vect{\Theta}}_{\rm LSTM})
\end{equation}

where $f_1(\cdot)$ represents the hidden state update function of the BiLSTM network 
and  specifically described by Equation.~(\ref{eq:1}) --- Equation.~(\ref{eq:2}).
It is necessary to point out that the complete 
output at each time point  are all a fusion feature obtained 
by weighting and summing the elements forward and backward according 
to the elements. Taking the time $t$ as an example, 
the complete output $\vect{h}_t$ can be calculated by the following formula:
\begin{equation}
    \vect{h}_i= \lambda_1 \cdot \vec{\vect{h}}_t \oplus \lambda_2 \cdot \cev{\vect{h}}_t
\end{equation}

\paragraph{Step 2} The output features extracted through the BiLSTM network 
 are feed into a FNN layer, The process is described as follows:
\begin{equation}
    \vect{o}_i =  f_2 \left(\vect{H}_i; \vect{\Theta}_{FC} \right)= g \left( \vect{W}_F \vect{H}_i  + \vect{b}_F \right)
\end{equation}

where we use ReLU function for $g$ in this work.

\paragraph{Step 3} The output feature $\vect{o}_i$ of the fully connected layer 
is finally input to a linear regression layer to calculate the host 
load prediction value $\hat{\vect{y}}_i=\left(\hat{\vect{x}}_{t+1},\hat{\vect{x}}_{t+2}, \cdots, \hat{\vect{x}}_{t+m}\right)$, 
the process can be described by the following formula:
\begin{equation}
    \hat{\vect{y}}_i=f_3 \left(\vect{H}_i;\vect{\Theta}_R \right)=\vect{W}_R \vect{o}_i
\end{equation}

where $\hat{\vect{y}}_i$ and $\vect{W}_R$  are the predicted value of the host load and 
the weight vector of the last linear regression layer.
 It is worth noting that, 
Depending on the performed tasks, these ‘prediction’ sequences can be either real
 host load values or mean load values over future time intervals.
  the input and output dimensions will 
also be different, respectively.

\subsection{Parameters of our model}\label{pom}

To train our network, we have used the backpropagation through time (BPTT) algorithm~\cite{58337}, 
which consists of repeated applications of the chain rule. Similar to 
the previous Google host load prediction work using LSTM, we `clip' the 
gradient before parameters update when it becomes to large to prevent 
exploding gradients. We have also used similar input layer size, 
hidden layer size, batch size, number of epochs, and learning rate, 
which is also annealed by 0.1 every 30 epochs.

In addition, due to the long length of our prediction method, 
we have used `truncated backpropagation through time' to 
reduce the cost of a single parameter update, similar to the 
previous LSTM work. The truncated length is kept at 39 
for the mean load prediction task; however, different from 
the previous LSTM work, this length is reduced to 26 for the 
real values prediction task due to long-term dependencies 
learning ability of LSTM-ED method. The parameters’ values 
of our LSTM-ED method are summarized in Table \ref{tbl:pom}.

\begin{table}[htbp]
    \caption{Parameters of our BiLSTM-based model.}
    \centering
    \label{tbl:pom}
    \begin{tabular}{p{20em} l}
        \toprule[1.5pt]
       Parameter & value    \\ \hline
       Input layer size  &  24/64  \\
       Hidden layer size &  128 \\
       Batch size &        128 \\
       Global gradient clipping norm &  5 \\
       Truncated length &  36/39 \\
       Epoch &      90   \\
       Dropout rate &   0.01 \\
       Early stop rate &   10\\
      \bottomrule[1.5pt]
      \end{tabular}
\end{table}

\section{Experiment results}\label{sec:4}

In order to evaluate our method, we have predicted both the actual load value and mean
 load value of the Google cluster workload traces~\cite{clusterdata:Wilkes2011}. Before training of 
 our method and all other benchmark methods, the input data are
  standardized by removing the mean and scaling to unit variance 
  to help with the convergence of gradient descent and performance of the methods.

\subsection{ Google load traces and baseline model }

The Google cluster 
records tracking data from more than 12,500 compute nodes for 29 days, with approximately 67,2074 jobs,
 more than 26 million resource usage data for more than 26 million tasks, and a measurement cycle of 5min 
 per record, each of which is divided into at least one task, each with corresponding scheduling constraints, 
 resource constraints, and detailed resource usage.

\begin{figure}[ht]
    \begin{subfigure}{0.22\textwidth}
      \centering
      \includegraphics[width=\linewidth]{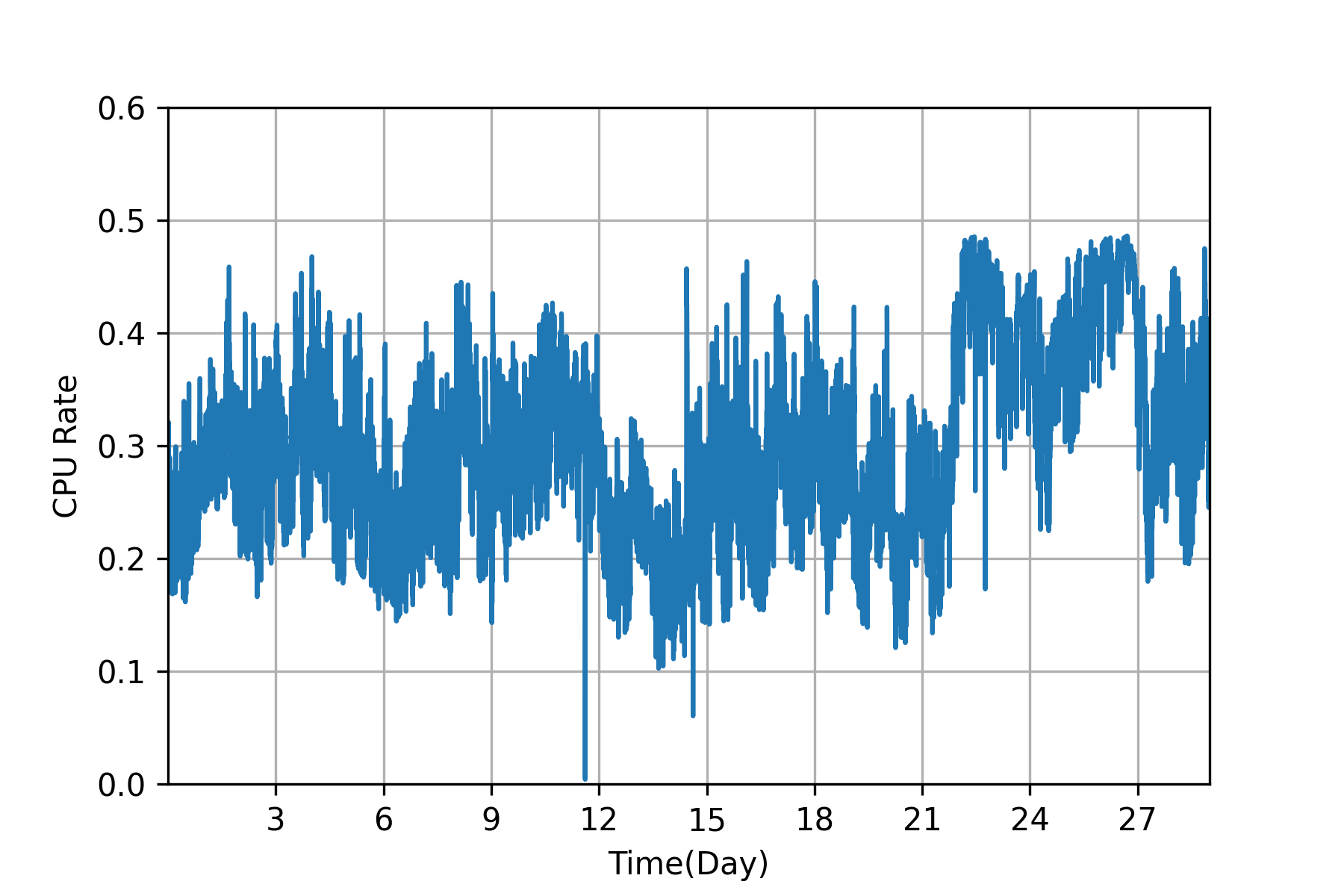}  
      \caption{Host load of the whole 29 days.}
      \label{fig:29d}
    \end{subfigure}
    \begin{subfigure}{.22\textwidth}
      \centering
      \includegraphics[width=\linewidth]{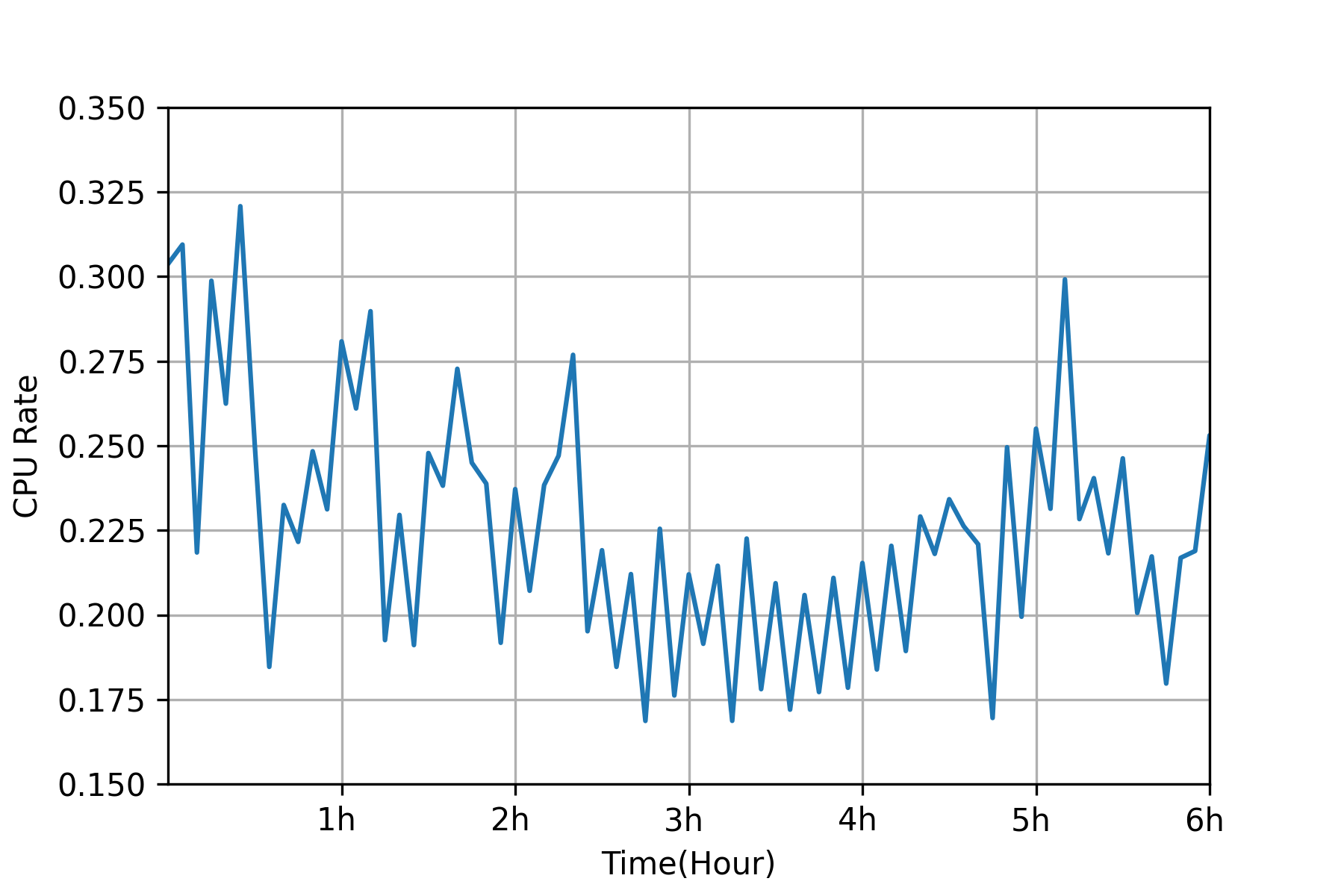}  
      \caption{Host load of the first 6 hours.}
      \label{fig:6hour}
    \end{subfigure}
    \caption{Host load of a single machine with id 5638349022 in Google load traces}
    \label{fig:fig}
\end{figure}

  In order to be able to obtain load data for each machine, our proposed method 
  uses the CPU of all 
  tasks that are running at the current time of each machine as the current host load, 
  and obtains the naturalized data through data preprocessing such as normalizetion
  to speed up the convergence of training. Figure~\ref{fig:29d} and Figure~\ref{fig:6hour} give the host load data 
  with machine number 563849022, where Figure~\ref{fig:29d}  shows the load data for 
  the full measurement period, 
  and Figure~\ref{fig:6hour} clearly depicts the fluctuation of of the host load in the first 6 hours.

In order to be able to follow up on the model of BiLSTM network sequence data modeling and 
host load regression analysis, our method uses a random sampling of 1000 machines 
as a data set to verify the validity of the BiLSTM network, first of all, 1000 machines 
29 days of tracking data divided into training set, validation set and test set, 
the first 20 days of tracking data as training data to calculate the parameter set of BiLSTM, 
The 21-26-day tracking data is used as a validation dataset to select model superparameters 
to avoid model over-fitting, and 27-29 days of tracking data as a test dataset to evaluate the proposed model.

\subsection{Accuracy metrics}

\paragraph{Mean load prediction} To make the results comparable with other methods, we first used LSTM model to 
predict the mean load. The metric named exponentially segmented pattern (ESP) was used to characterize the host load
fluctuation over consecutive time intervals whose lengths increase exponentially.

The mean segment squared error  (MSSE) defined as follows was applied to quantity the 
performance of mean load prediction:

\begin{equation}
    \textrm{MSSE}(s)= \frac{1}{s} \sum_{i=1}^n s_i \left(l_i-L_i \right)^2
\end{equation}

where $s_1 = b, s_i = b \cdot 2 ^{i-2}, s = \sum_{i=1}^n s_i$, 
$b$ is called baseline segment, which is set 
to 5 min, similar to previous LSTM work~\cite{song2018host} and LSTM-ED work~\cite{nguyen2019host}; 
$l_i$ is the predicted mean value. $L_i$ is the  true value, $n$ is the number of segments in
the consecutive prediction interval. In practice, the load pattern is converted 
from single load interval.

\paragraph{Actual load prediction} 

Because the length of the segment  shows an exponential growth trend across time in the mean
load prediction task, 
so as the length of the segment increases, 
the average load value based on the segmentation mode can no longer 
fully capture the actual fluctuation of  the load, so we also specifically 
considers the actual load prediction task.
The forward load of the model can be 
used to obtain the actual load sequence for a period of time in the future. 
It is similar to the actual load prediction model of others~\cite{peng2018multi,song2018host,nguyen2019host}, 
and the evaluation method based on the Mean Squared Error (MSE) is defined to
 measure the model:

\begin{equation}
    \textrm{MSE}= \frac{1}{N} \sum_{i=1}^N  \left(\hat{y}_i-y_i \right)^2
\end{equation}

where  $N$ is the length of the prediction step, 
and $y_i$ and $\hat{y}_i$ are the actual and forecast values, 
the interval length between two consecutive host load values is set 
to 5 min for the Google dataset, respectively.

\subsection{Analysis of different prediction lengths}

Table~\ref{tbl:1} shows the impact of the five different prediction steps 
on the mean load prediction. From the table, it can be clearly 
seen that the proposed method has a good performance in the 
short-term mean load forecasting task. As the
prediction steps increases, the MSSE
also increases, which can be explained mainly from two aspects.
Firstly, the BiLSTM network extracts effective temporal features by learning past
 and future load changes, and ultimately produces good short-term 
 load prediction results. Secondly, when the number of prediction 
 steps increases, 
although the input sequence data already contains more temporal information, 
the BiLSTM network still has deficiencies in capturing long-term dependencies.

\begin{table}[htbp]
    \caption{Impact of prediction length on mean load forecast.}
    \centering
    \label{tbl:1}
    \begin{tabular}{P{10em} c c}
        \toprule[1.5pt]
       Prediction length &  MSSE  &  Training Time (s)  \\ \hline
0.7h & 0.00055 & 2.3745 \\
1.3h & 0.00059 & 2.3694 \\
2.7h & 0.00109  & 2.4495 \\
5.3h & 0.00139  & 2.4189 \\
10.7h & 0.00151 & 2.4331 \\
      \bottomrule[1.5pt]
      \end{tabular}
\end{table}


we also studies the impact of forecasting step size on actual load forecasting.
 However, considering the actual load forecasting step size is different from the 
 mean load prediction segmentation mode, the forecasting step size is designed
  with a multiple of 0.5h. From Table~\ref{tbl:2}, 
it can be seen from the MSE and the training time that actual 
load prediction is more challenging than the mean load prediction task, 
and from Table~\ref{tbl:2}, it can be further seen that as the number of prediction 
steps increases, the MSE shows that upward trend.

\begin{table}[htbp]
    \caption{Impact of prediction length on actual load forecast.}
    \centering
    \label{tbl:2}
    \begin{tabular}{P{12em} c c}
        \toprule[1.5pt]
       Prediction length~(5 min) &  MSE  &  Training Time (s)  \\ \hline
       6 & 0.001162 & 5.087 \\
       12 & 0.001784 & 5.175 \\
       18 & 0.002108 & 5.222 \\
       24 & 0.002470 & 5.330 \\
       30 & 0.002679 & 5.551 \\
       36 & 0.002995 & 5.744 \\
      \bottomrule[1.5pt]
      \end{tabular}
\end{table}

\subsection{Mean load prediction}

\begin{figure}[htbp]
    \centerline{\includegraphics[width=0.4\textwidth]{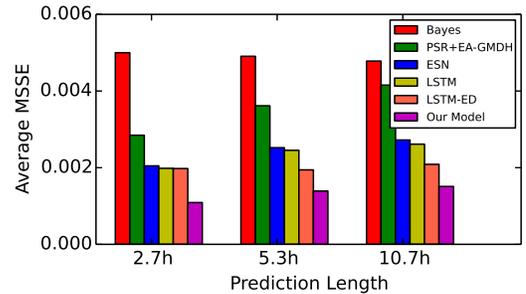}}
    \caption{Comparison between our method with five others for consecutive intervals}
    \label{fig:msse}
    \end{figure}

As you can see from the Figure~\ref{fig:msse},
Compared with the methods based on Bayes~\cite{di2012host}, 
PSR+EA-GMDH~\cite{yang2013host}, AutoEncoder+ESN~\cite{yang2015multi}, LSTM~\cite{song2018host}, etc.,
   our proposed methods are significantly better than other methods in 
the mean load prediction task. 
Compared with the state-of-art methods,
 such as LSTM-ED~\cite{nguyen2019host}, the prediction performance 
of our method  also has advantages. 
Although the prediction 
performance based on the LSTM-ED method with the increase of the prediction length 
is relatively stable, this is because the method uses the encoder-decoder structure 
and introduces the context vector to save the feature information for a long time, 
which can enable the model to obtain longer temporal information. Compared with the 
method based on LSTM-ED, the method based on BiLSTM network mentioned 
 has achieved better prediction performance for the mean load prediction 
task on the Google cluster trace dataset.

\begin{figure}[htbp]
    \centerline{\includegraphics[width=0.4\textwidth]{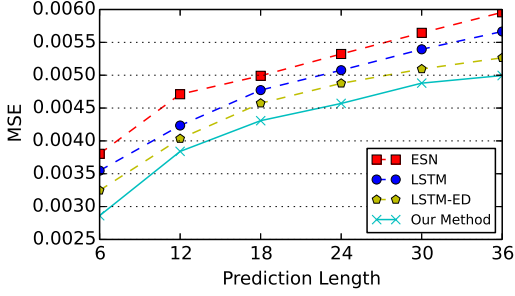}}
    \caption{MSE result comparison among four methods with interval length of 5 min.}
    \label{fig:mse}
    \end{figure}

Figure~\ref{fig:mse} shows the comparison of host load prediction performance 
between our proposed model and AutoEncoder+ESN~\cite{yang2014new}, LSTM~\cite{song2018host}, 
and LSTM-ED~\cite{nguyen2019host} under 
different prediction lengths.Figure~\ref{fig:mse} clearly shows that the prediction accuracy of 
the model proposed is better than the other three models at all prediction 
lengths. Different from the AutoEncoder+ESN model, different from the LSTM and LSTM-ED models, 
the BiLSTM model mentioned  can not only model the temporal dependency 
from the historical information of the host load, but also extract the feature 
information from the future host load changes, Compared with LSTM and 
LSTM-ED, BiLSTM-based model can obtain more powerful nonlinear generalization ability.

    \subsection{Actual load prediction}

\begin{figure}[htbp]
        \centerline{\includegraphics[width=0.48\textwidth]{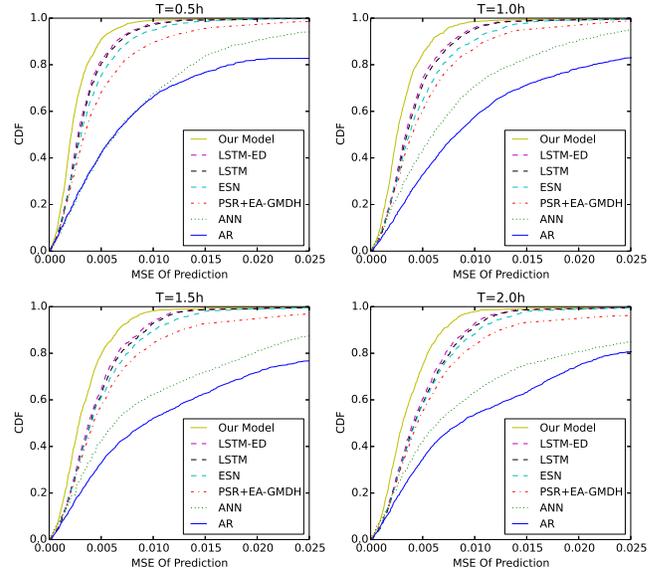}}
        \caption{Cumulative Distribution Function (CDF) of MSE results among seven different methods.}
        \label{fig:cdf}
    \end{figure}

    In order to better show the actual load prediction performance comparison, 
    our method also considers the cumulative distribution function (CDF) of all comparison models MSE, 
    that is, the probability distribution of all data less than or equal to the current MSE value, 
    CDF is accurate for the analysis modelHarmony and stability are of great help. 
    In the field of host load prediction, many studies~\cite{song2018host,yang2014new,nguyen2019host}
    have adopted MSE's CDF as 
    an indicator to evaluate the accuracy of the model. In particular, the MSE's CDF graph 
    can clearly depict the distribution of different models' MSE.
    Figure~\ref{fig:cdf} depicts the CDF 
    of the MSE results of seven different prediction methods. Each data point in the CDF
    curve describes the proportion of all MSE values that are less than or equal to the
    horizontal coordinate of the point (that is, the value corresponding to the vertical
     coordinate of the point), it is worth noting that the closer the CDF curve is to the 
     vertical axis, the higher the accuracy of the corresponding model.
    It can be seen from Figures~\ref{fig:cdf} that the AR-based model~\cite{4354138} and 
    the ANN-based model~\cite{5160878} are different from the CDF curve of other models. 
    Under different predicted lengths, the ordinate value corresponding to 0.025 on 
    the horizontal axis of the CDF Significantly less than 1, that is, the proportion 
    of all data points whose MSE is less than or equal to 0.025 is significantly less 
    than 1.Therefore, they have poor accuracy, leading to low accuracy for two main 
    reasons: (1) their non-linear generalization ability is poor and they cannot
     complete the multi-step prediction task in advance; (2) they cannot effectively 
     use the historical host load fluctuation information. 
     In addition, you can see that the AutoEncoder+ESN,
      LSTM, LSTM-ED, and our proposed method,
all MSE values on 1000 machines are basically less than 0.025. This is mainly because
 they are all RNN-based models. The load data is modeled, and it can be seen that as the predicted 
 length increases, the CDF curve of MSE also shifts to the right.
 The corresponding map 3-10 
 depicts the MSE distribution of the different prediction methods under the load data of
  all test hosts.

\begin{figure}[htbp]
    \centerline{\includegraphics[width=0.48\textwidth]{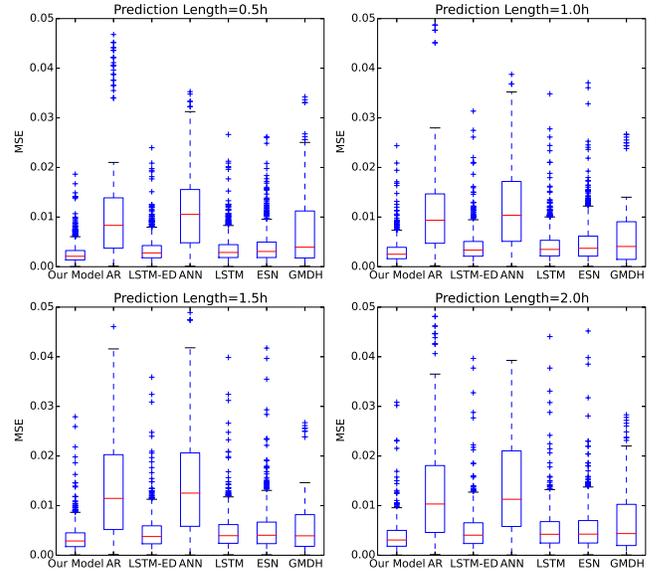}}
    \caption{Boxplot of MSE results among seven different methods.}
    \label{fig:boxplot}
\end{figure}

Figure~\ref{fig:boxplot} is expressed in the form of a box plot. 
The box plot can clearly see all the prediction errors under different
 prediction lengths.The median is smaller than the other 6 prediction models.
  Therefore, under all prediction lengths, our proposed method 
  has better prediction accuracy than the other 6 models.

\subsection{Prediction results}

\begin{figure*}[htbp]
    \centerline{\includegraphics[width=\textwidth]{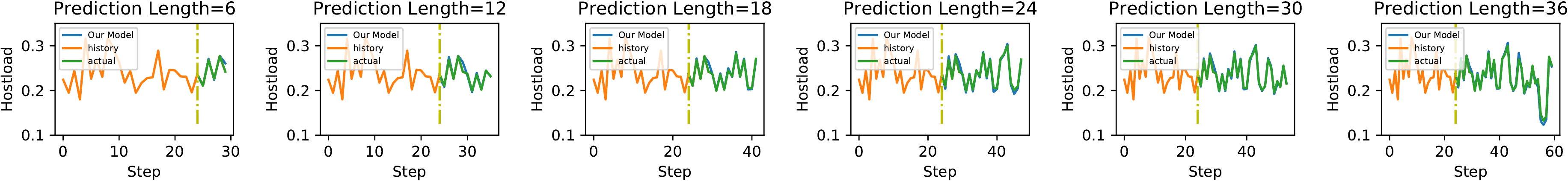}}
    \caption{The actual cpu load prediction results with different prediction lengths in Google cluster data.}
    \label{fig:cpu}
    \end{figure*}

\begin{figure*}[htbp]
        \centerline{\includegraphics[width=\textwidth]{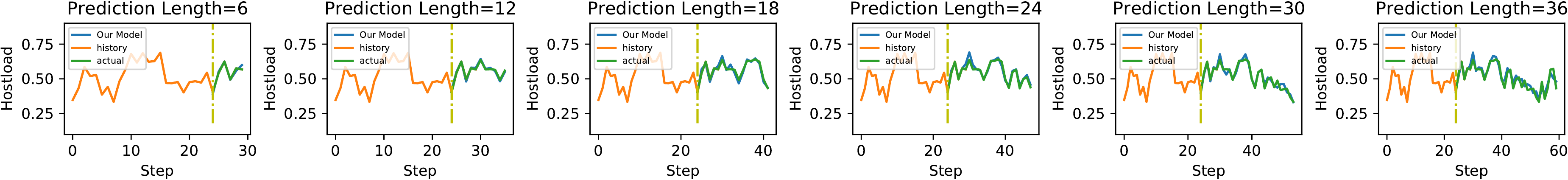}}
        \caption{The actual memory load prediction results with different prediction lengths in Google cluster data.}
        \label{fig:mem}
        \end{figure*}

In order to give an insight into the predictions, two results are present in Figure.~\ref{fig:cpu}
 and Figure.~\ref{fig:mem} which both have poor autocorrelation and drastic fluctuates. 
 One is the host cpu load on the Google cluster data  in different 
 prediction length,
  The other is the host memory utilization prediction results with 
 different prediction length which \verb|machine_id|=5411731657 in Google cluster data. In this case, 
 Our method can give  statisfactory performances  and perfect predictions 
 for six different prediction lengths.

\section{Conclusion}\label{sec:5}
We have proposed a method based on Bi-directional Long  Short-Term Memory  (BiLSTM)
for host load prediction. The proposed method can effectively perform  the mean load prediction task
and the actual load prediction task using Google load trace data. 
Different from other deep learning based methods, our proposed method  
can simultaneously consider the historical information and future information of the host load. 
Because our proposed model  is limited with size, 
the discarding method and the early stopping method are used to alleviate the overfitting 
problem in the model training process.
 The effect of further prediction length on the prediction 
performance of the proposed method is analyzed through numerical examples, 
and the effectiveness of the proposed method in short-term load and 
long-term load prediction tasks is further verified through comparative experiments. 
Finally, the prediction performance of our proposed  method  is compared 
with the state-of-art methods, and the results also verify 
the effectiveness of our proposed method.
In addition, our proposed method
 can concisely and effectively implement a series of tasks such as Google cloud trace
   data collection and host load prediction. Therefore, the research 
    can provide new solutions and ideas for end-to-end host load prediction in  cloud computing environment.

\section*{Acknowledgment}


This work was partially supported by  Grant No. S2019ZPYFB0553 from the Primary Research and Developement Plan of Jiangxi Province and 
Grant No. 2016YFC14017062016 from the National Key Research and Development Program of China.





\bibliographystyle{IEEEtran}
\bibliography{ms}

\vspace{12pt}
\begin{IEEEbiography}
    [{\includegraphics[width=1in,height=1.25in,clip,keepaspectratio]{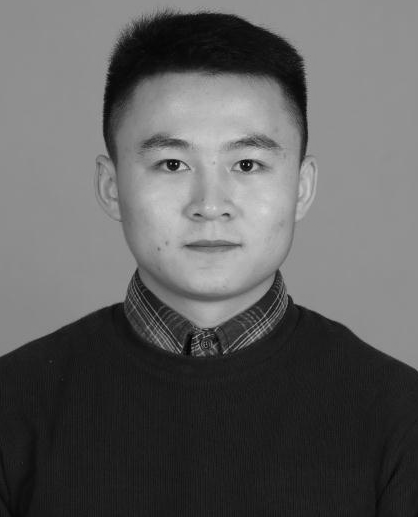}}]{Hengheng Shen}
    is received Bachelor’s degree in Computer Science and 
    Technology from Anyang Institute of Technology in 2017, Anyang, China. 
    Master’s degrees in Institute of Computing Technology, Chinese Academy,
     Beijing, China. His research interests are in Cloud Computing and Machine Learning.
\end{IEEEbiography}

\begin{IEEEbiography}
    [{\includegraphics[width=1in,height=1.25in,clip,keepaspectratio]{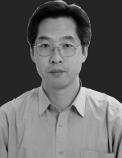}}]{Xuehai Hong}
    is received the Postdoctoral in department of computer science from Beijing University in 2003.
     He is an professor level senior engineer in the 
     Institute of Computing Technology, Chinese Academy. 
     His research interests are in high perfermance computing, 
     cloud computing and bid data.
\end{IEEEbiography}
\end{document}